Paper Type: Research Paper

# Application of Accelerated Life Testing in Human Reliability Analysis

**Rasoul Jamshidi**[1],*  , **Mohammad Ebrahim Sadeghi**[2]

[1] Faculty of Engineering School, Damghan University, Semnan, Iran; r.jamshidi@du.ac.ir.
[2] Faculty of Management, Department of Industrial Management, University of Tehran, Tehran, Iran; sadeqi.m.e@gmail.com.

## Abstract

As manufacturers and technologies become more complicated, manufacturing errors such as machine failure and human error have also been considered more over the past. Since machines and humans are not error-proof, managing the machines and human errors is a significant challenge in manufacturing systems. There are numerous methods for investigating human errors, fatigue, and reliability that categorized under Human Reliability Analysis (HRA) methods. HRA methods use some qualitative factors named Performance Shaping Factors (PSFs) to estimate Human Error Probability (HEP). Since the PSFs can be considered as the acceleration factors in Accelerated Life Test (ALT). We developed a method for Accelerated Human Fatigue Test (AHFT) to calculate human fatigue, according to fatigue rate and other effective factors. The proposed method reduces the time and cost of human fatigue calculation. AHFT first extracts the important factors affecting human fatigue using Principal Component Analysis (PCA) and then uses the accelerated test to calculate the effect of PSFs on human fatigue. The proposed method has been applied to a real case, and the provided results show that human fatigue can be calculated more effectively using the proposed method.

## 1 | Introduction

Today we see comprehensive automation and consequently the reduction of the human role in most manufacturing systems. Although these efforts reduce human error and increase product quality, humans still have an important effect on quality in manufacturing systems. Human factors (i.e., fatigue, recovery) can have a significant impact on the performance of the overall system [1].

In order to prevent or reduce human fatigue and disorders, human reliability and its related issue are studied and investigated in manufacturing contexts such as design work process, equipment design, and ergonomic consideration. Battini et al. [2] proposed that human disorder is one of the important causes of health problems in workers. Different methods have been proposed to mitigate the human disorders in manufacturing systems.

Some methods highlighted the workplace design and environmental conditions [3] and [4], and other methods investigated the work redesigning and redefining [5]. The common point of all these methods is that some important factors such as human reliability and fatigue should be measured in order to evaluate the severity of the effect on quality and cost. Other studies investigated the human reliability calculation methods such as the first and second generation of HRA and introduced new concepts such as fatigue in HRA.

HRA methods are time consuming, and new methods must be introduced to reduce the implementation cost and time. In this paper, we propose a hybrid method that uses the advantage of ALT as a reliability analysis model to predict the fatigue value. The effective parameters on human fatigue are also extracted by PCA. This hybrid method reduces the time and costs in HRA.

## 2 | Literature Review

There are several sources of human error; one of them is fatigue that stems from other factors; in fact, every loss caused by psychological or physiological effort is named fatigue that has several symptoms. Lack of energy, physical exertion [6], lack of motivation [7], and sleepiness [8] are the symptoms of fatigue. Human fatigue commonly disrupts human performance and leads to undesirable costs to the manufacturer such as quality cost, maintenance cost, etc. With unfavorable effects on judgment, efficiency, and productivity [9].

Many researchers investigated fatigue in qualitative and quantitative categories. In the first category, they studied the effective factors in human fatigue, such as the type of work, environmental condition, and human body specification [10] and [11].

The second category investigated the methods to calculate the fatigue value due to qualitative factors. Also, some papers in this category proposed different methods for human fatigue recovery. [12]-[14] proposed several methods to calculate the fatigue based on some factors such as Maximum Holding Time (MHT) and Maximum Endurance Time (MET). Imbeau and Farbos [15] proposed the static fatigue analysis through the concept of MET, which defines the maximum time a muscle can sustain a load. Ma et al. [16] studied the influence of external load on human fatigue in a real-time situation. Fruggiero et al. [17] investigated the imposed uncertainty due to human error. Peternel et al. [18] proposed two fatigue management protocols to calculate and reduce human fatigue. Li et al. [19] introduced the fatigue causal network for fatigue management.

Since the first proposed methods for fatigue measuring only consider the work type and some human specifications, PSFs have been proposed to increase the accuracy of fatigue calculation.

The term PSF encompasses the various factors that affect human performance and can change the HEP. Although many methods use PSFs, there is not a standard set of PSFs that are used for most HRA methods [20]. There are about 60 PSFs, with varying degrees of overlap [21]. Boring studied the important PSFs and proposed 8 PSFs that are considered in common HRA methods [22], such as stress, complexity, ergonomics and etc. Also, the PSFs affect human fatigue/recovery and should be considered in human fatigue quantification. Rasmussen and Laumann [23] proposed a method to evaluate fatigue as a performance shaping factor in the Petro-HRA method.

Some researchers studied human fatigue from the perspective of reliability and believed that human fatigue could be investigated using reliability methods [24]. In fact, they considered human as a machine and proposed some reliability and maintenance policy to mitigate the human failure rate. For example Mahdavi et al. [25] proposed a mathematical model for a dynamic cellular manufacturing systems whit human resources. Cappadonna et al. [26] addressed the machine scheduling problem with limited human resources. Taylor [27] studied the human as an important resource in maintenance actions. The main problem to study human fatigue by machines reliability methods is that the PSFs have not been considered by reliability methods. Griffith and Mahadevan [28] investigated the effect of fatigue on human reliability.

ALT can be used to consider the PSFs in reliability assessment methods and evaluate human fatigue. ALT is the process of testing an element subjecting it to environmental conditions such as stress, temperatures [29] and [30]. PSFs can be considered as an Accelerated Factor (AF) to calculate human fatigue. In fact, ALT reduces the time and cost of human fatigue calculation.

To the best of author's knowledge, there is no a research in which studied human fatigue using ALT and compared its result with common fatigue models. In this paper, we propose an Accelerated Life Test (ALT) to calculate the human fatigue value to consider the effect of PSFs in fatigue calculation and make fatigue value closer to reality. To find the most effective PSFs we utilize Principal Component Analysis (PCA) since the ALT needs a limited number of factors. Also, a real case will be investigated to examine the effectiveness of the proposed methods. In other words, we select the most effective PSFs with PCA and then calculate the fatigue by ALT. this method reduces the time and decreases the cost of human fatigue calculation. Also, eliminating unnecessary PSFs reduces the data gathering process without significant prediction error.

## 3 | Research Methodology

In the proposed method, we first select the most effective PSFs on human fatigue using the PCA method. By PCA, we can limit the PSFs and consider the important PSFs to facilitate the data gathering. After selecting the PSFs, we use the ALT to evaluate the effect of PSFs on human fatigue. Using ALT, we can calculate the fatigue according to the PSfs value with a non-significant error.

### 3.1 | Principal Component Analysis (PCA)

PCA is an unsupervised exploratory method for feature extraction. This method combines several input variables in a specific way to drop the "least important" variables and retains the most valuable parts of all of the variables. PCA produces orthogonal components by decomposing the initial input variables matrix [31].

PCA is mostly used for making predictive models. PCA can be done based on the covariance matrix as well as the correlation matrix, and data matrix using eigenvalues and eigenvectors. The data matrix should be normalized until having zero mean and unit variance [32]. The results of a PCA are usually studied in terms of component scores, or factor scores. The steps of PCA implementation for some stochastic vector *X* is as follows:

I. Consider we have a matrix X with n rows and p+1 columns, where there is one column corresponding to the dependent variable (usually denoted Y) and p columns corresponding to each of independent variables.
II. For each column, subtract the mean of that column from each entry.
III. Make the matrix Z by standardizing each column of X to make sure each column has mean zero and standard deviation 1.
IV. Calculate the $Z^T Z$.
V. Calculate the eigenvectors and eigenvalue for $Z^T Z$.
VI. Take the eigenvalues $\lambda_1, \lambda_2, \ldots$ and sort them from largest to smallest.
VII. Calculate the proportion of variance explained by *Eq. (1)* for each input variable.

$$\text{proportion of variance explained}_i = \frac{\lambda_i}{\sum_{i=1}^{n} \lambda_i}. \qquad (1)$$

VIII. For each variable, pick a threshold and add input variables until the cumulative proportion of variance hits that threshold.

IX. Select the last input variable that hits the threshold and its previous variable as effective inputs variable on Y.

Using PCA in human fatigue analysis, we can eliminate the probable dependency between PSFs and obtain the main effective PSFs on human fatigue. On the other hand, by reducing the size of input variables, there is no need to spend more time and cost for PSFs data gathering.

### 3.2 | Accelerated Life Testing (ALT)

All consumers are interested to know the lifetime knowledge of products. Using this knowledge, manufacturers can estimate their cost of production by knowing the failure mode of their machines [33] and [34], and end customers can predict the maintenance cost of their appliances. The lifetime of most machines is long and the manufacturers cannot implement common lifetime experiments to ensure the failure rate of machines. To overcome this issue ALT has been proposed.

ALT identifies the load stress levels of a system (or machines or components) and studies the effects of increasing this stress during the life test [35]. In other words, during the ALT, the machine works with a stress higher than its common levels of loads or different environmental factors in order to accelerate the failure occurrence cycle [36]. ALT reduces the testing time to estimate behavioral characteristics such as failure rate, and lifetime for a machine in normal conditions. In ALT, all stress levels are not examined, and the test is conducted in some stress levels, and extrapolation is used to estimate the life distribution at the desired conditions. Since extrapolation methods have statistical errors, several ALT have been proposed by researchers to eliminate the extrapolation errors.

To use ALT, the failure distribution should be identified, and the AF should be defined to propose the ALT distribution function [36]. Three well-known distributions in ALT are Weibull, Exponential, and Log-normal distribution [37]. Weibull is most used among these three distributions [38].

The acceleration relationship selection is important in ALT; this relationship is selected according to the types and numbers of AF. For example, if the temperature is considered as AF, the Arrhenius is an appropriate relation. Eyring Inverse Power Law (IPL), Temperature-Humidity, Temperature Non-Thermal Relationship are the most used relation in ALT.

In this paper, we consider the Performance Shaping Factors (PSFs) as AF, and select the proper acceleration relationship according to the number of effective PSFs obtained by PCA methods. In fact, we aim to propose a quantitative method for human fatigue and reliability calculation instead of common qualitative human fatigue measurement methods.

### 3.3 | Human Fatigue

Fatigue refers to the issues that arise from excessive working time, poorly designed shift patterns, inappropriate tools, and poor ergonomics. Fatigue is usually considered to be a decrease in mental or physical performance of human that results from long exertion, sleep loss, or disruption of the body's internal clock. It is also related to workload, in that workers are more easily fatigued if their work is machine-paced, complex or monotonous. Fatigue leads to a reduction in human's workforce and consequently to increase the Human Error Probability (HEP).

The physical fatigue involvement in HEP has been investigated by some researchers. Studies on muscle fatigue have been going on since the late 1800s [39] and [40], and research has mainly focused on the fatigue rate in terms. Xia and Law [41] presented a three-state model to assess human fatigue. Myszewski [42] proposed a curve-based model on error rate to show that human error increases as fatigue increases over time. Michalos et al. [43] proposed a scoring method for physical fatigue using the fatigue model of [16]. They also provided a method to calculate the corresponding error rate based on the work of [9]. Jamshidi and Seyyed Esfahani [44] proposed a model to assess the fatigue of human resources in production systems. Since fatigue is a wide-ranging term, it cannot be directly evaluated. Fatigue must be inferred from its related factors, such as excessive sleepiness, reduced physical and failure rate of human resources [45]. For physical fatigue, fatigue can be measured using physical factors such as heart rate [46] and force. One of the most popular fatigue relations has been proposed by [47]. He proposed that fatigue value can be calculated using *Eq. (2)*.

$$f(t) = 1 - e^{-\lambda_f \cdot t}. \qquad (2)$$

Where f(t) is the fatigue accumulated by time t, λf is the human fatigue rate. The fatigue rate represents the rate at which the fatigue occurs.

The failure rate is affected by many factors such as human skill, training, work type, equipment, and environmental condition such as temperature, light, vibration, etc. therefore, we can assign a unique failure rate to a human in a specific job implementation. Finding the human fatigue rate in each environment requires several experiment implementations with different conditions. These different conditions have been known as PSFs. In this paper, we study the historical data on human failure and select the most effective PSFs on human fatigue by PCA method; then the most effective PSFs are considered as AF to propose the accelerated fatigue relation that considered the environmental conditions in production systems. The accelerated fatigue relation will be presented in the next section.

### 3.4 | PSFs Selection by PCA

PSFs can enhance or degrade human performance and provide a basis for considering potential influences on human performance and considering them in the quantification of HEP. There are several categories of PSFs such as direct/ indirect, internal/external, etc. The most popular PSFs are shown in *Table 1*.

Table 1. Most popular PSFs.

| No. | PSF Title           | No. | PSF Title               |
|-----|---------------------|-----|-------------------------|
| 1   | Training & Experience | 9   | Equipment Accessibility |
| 2   | Available Time      | 10  | Fitness for Duty        |
| 3   | Team/Crew Dynamics  | 11  | Instrument Availability |
| 4   | Environment         | 12  | Workload/Stress         |
| 5   | Communications      | 13  | Ergonomics/HIS          |
| 6   | Communications      | 14  | Need for Special Tools  |
| 7   | Complexity          | 15  | Realistic Accidents     |
| 8   | Available Staffing  |     |                         |

Considering the PSFs proposed in *Table 1*, it's hard to analyze all the 15 numbers of PSFs in manufacturing systems, to overcome this issue, researchers tried to reduce the effective PSFs. Boring studied the important PSFs and proposed eight PSFs that are considered as the most effective PSFs in HRA methods [22] Available Time, Stress, Complexity, Experience and Training, Procedures,

Ergonomics, Fitness for Duty, Work Process are the eight effective PSFs. Each PSF has a predefined level and each level has a specific value [48] and [49]. For example, the values of stress are shown in *Table 2*.

Table 2. The value of PSF level (Stress).

| PSF level (Stress) | Multipliers Action | Multipliers Diagnosis |
|---|---|---|
| Extreme | 5 | 5 |
| High | 2 | 2 |
| Nominal | 1 | 1 |
| Insufficient information | nominal | nominal |
| Highly complex | 5 | 5 |
| Moderately complex | 2 | 2 |

The multiplier values were attributed by analysts on the basis of several studies. These multipliers should be standardized, if we face an abnormal manufacturing system. In this paper, we consider these 8 PSFs and investigate the human fatigue, regarding these PSFs value and then extract the most effective PSFs using PCA. In order to find the effective PSFs, we investigated 15 historical data for a lathing workshop that presented in *Table 3*, the time for fatigue measurement is equal to 1 hour, each time unit are considered to be 1 hour.

Table 3. The PSFs value for each instance.

| PSFs | Available Time | Stress | Complexity | Experience And Training | Procedures | Ergonomics | Fitness For Duty | Work Process | Fatigue |
|---|---|---|---|---|---|---|---|---|---|
| Ins 1 | 0.1 | 2 | 5 | 3 | 20 | 0.5 | 5 | 0.5 | 0.130 |
| Ins2 | 10 | 2 | 2 | 0.5 | 1 | 10 | 1 | 0.5 | 0.110 |
| Ins 3 | 10 | 1 | 2 | 1 | 50 | 1 | 5 | 5 | 0.126 |
| Ins 4 | 0.1 | 1 | 1 | 1 | 5 | 0.5 | 1 | 0.5 | 0.035 |
| Ins 5 | 10 | 2 | 5 | 3 | 50 | 1 | 5 | 5 | 0.165 |
| Ins 6 | 0.1 | 2 | 5 | 0.5 | 1 | 1 | 5 | 1 | 0.078 |
| Ins 7 | 0.01 | 1 | 1 | 0.5 | 1 | 1 | 1 | 5 | 0.027 |
| Ins 8 | 0.01 | 5 | 2 | 1 | 1 | 1 | 1 | 5 | 0.086 |
| Ins 9 | 0.01 | 5 | 2 | 0.5 | 1 | 10 | 5 | 5 | 0.138 |
| Ins 10 | 10 | 2 | 1 | 3 | 1 | 10 | 5 | 5 | 0.150 |
| Ins 11 | 1 | 1 | 5 | 0.5 | 1 | 10 | 1 | 5 | 0.094 |
| Ins 12 | 0.1 | 5 | 5 | 3 | 50 | 10 | 1 | 1 | 0.157 |
| Ins 13 | 0.01 | 5 | 5 | 0.5 | 20 | 10 | 5 | 0.5 | 0.142 |
| Ins 14 | 0.1 | 5 | 2 | 3 | 5 | 0.5 | 5 | 1 | 0.126 |
| Ins 15 | 0.1 | 5 | 2 | 3 | 5 | 0.5 | 5 | 1 | 0.134 |

To reduce the number of PSFs to make the data easier to analyze, the PCA method will be implemented. The results of PCA method are shown in *Table 4*.

Table 4. Eigen analysis of the Correlation Matrix.

| | | | | | | | | | |
|---|---|---|---|---|---|---|---|---|---|
| **Eigenvalue** | 2.7430 | 1.7996 | 1.3479 | 1.1389 | 0.7193 | 0.6709 | 0.3977 | 0.1681 | 0.0145 |
| **Proportion** | 0.305 | 0.200 | 0.150 | 0.127 | 0.080 | 0.075 | 0.044 | 0.019 | 0.002 |
| **Cumulative** | 0.305 | 0.505 | 0.655 | 0.781 | 0.861 | 0.936 | 0.980 | 0.998 | 1.000 |

The first three PSFs (available time, stress, and complexity) explain 65.5% of the variation in the data of *Table 3*. Therefore, these three PSFs can be used to analyze the fatigue value. The impact of each PSF is shown in *Fig 1*.

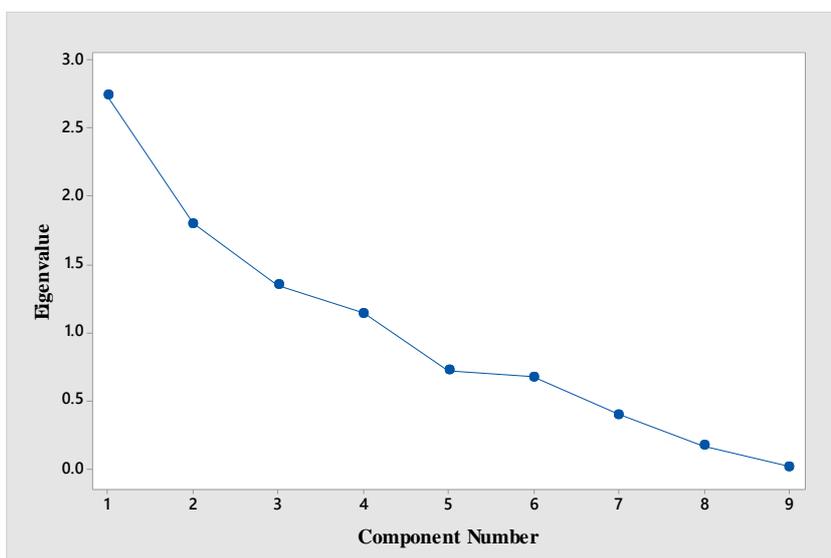

**Fig. 1. The eigenvalue for each PSF.**

On the other hand, the third PSF (complexity) has a positive correlation with the second PSF (stress), and we can find if the work complexity increases, then the worker stress also increases. Regarding this issue, we select the Available Time and stress as the two effective PSFs on human fatigue. The proposed data in *Table 5* also confirmed these results. As it could be seen, the available time and stress have the most correlation with fatigue. The fatigue correlation value with available time and stress is 0.276 and 0.313, respectively. *Fig. 2.* shows the correlation of these 8 PSFs with the most effective PSFs.

Table 5. Eigenvectors for each PSF.

| Variable | PC1 | PC2 | PC3 | PC4 | PC5 | PC6 | PC7 | PC8 | PC9 |
|---|---|---|---|---|---|---|---|---|---|
| Available Time | 0.165 | -0.641 | -0.043 | -0.090 | 0.214 | -0.342 | 0.197 | -0.523 | 0.276 |
| Stress | 0.244 | 0.489 | -0.163 | -0.421 | 0.042 | 0.370 | 0.365 | -0.358 | 0.313 |
| Complexity | 0.298 | 0.180 | -0.219 | 0.651 | -0.339 | 0.024 | -0.315 | -0.403 | 0.173 |
| Experience And Training | 0.421 | 0.041 | 0.362 | -0.143 | 0.462 | 0.100 | -0.610 | 0.103 | 0.251 |
| Procedures | 0.402 | -0.234 | 0.064 | 0.439 | 0.165 | 0.362 | 0.497 | 0.411 | 0.115 |
| Ergonomics | 0.084 | -0.023 | -0.809 | -0.123 | 0.126 | -0.209 | -0.156 | 0.411 | 0.265 |
| Fitness For Duty | 0.383 | -0.001 | 0.279 | -0.260 | -0.653 | -0.370 | 0.083 | 0.279 | 0.245 |
| Work Process | -0.076 | -0.509 | -0.125 | -0.252 | -0.393 | 0.650 | -0.279 | -0.017 | 0.041 |
| Fatigue | 0.571 | -0.048 | -0.200 | -0.178 | 0.008 | -0.035 | -0.019 | -0.100 | -0.767 |

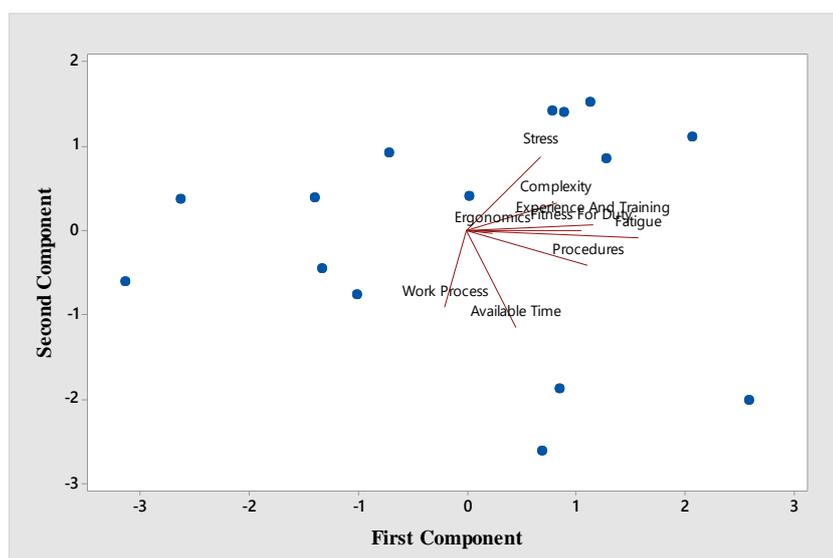

**Fig. 2. The relation between PSFs.**

### 3.5 | Acceleration Human Fatigue Test (AHFT)

The ALT methods have been proposed for one or more AF to estimate the reliability and fatigue. In most practical applications, fatigue is a function of more than one variable (stress types). In these cases, the General Log-Linear (GLL) can be used. GLL describes a life characteristic as a function of a vector of variables such as X=(X1,X2,...Xn), the GLL relation is as follows:

$$L(x) = e^{\alpha_0 + \sum_{j=1}^{n} \alpha_j X_j}. \qquad (3)$$

Where α0 and αj are the relation parameters that should be estimated and X is the AF vectors. This relationship can be further modified through the use of transformations and can be reduced to other ALT methods such as Arrhenius and IPL. Regarding this fact, we can use IPL method instead of GLL by using an appropriate transformation. To implement the ALT for human fatigue we use the GLL method with some transformation and consider the available time and stress as the AFs. Since we have 2 AFs we should estimate the α0, α1, α2 as the GLL parameters so we can predict the fatigue value according to stress and available time for the future dataset. The results of GLL implementation have been shown in *Table 6*, the confidence level is equal to 99%.

Table 6. Regression table.

| Predictor | Coef | Standard Error | Z | P | Lower 99.0% Normal CI | Upper 99.0% Normal CI |
|---|---|---|---|---|---|---|
| **Intercept** | 36.60 | 15.72 | 2.33 | 0.02 | (3.90) | 77.09 |
| **Available time** | 0.05 | 0.02 | 2.38 | 0.02 | (0.00) | 0.09 |
| **Stress** | (10,723.90) | 4,344.03 | (2.47) | 0.01 | (21,913.40) | 465.56 |
| **Shape** | 3.64 | 0.86 | | | 1.97 | 6.71 |

Regarding the result presented in *Table 6*, we can estimate the fatigue of human resources with different available times and stress levels. For example, the estimate of fatigue for available time= 0.1 and stress= 5 is shown in *Table 7*.

Table 7. The estimation of fatigue with PSFs value.

| Available time | Stress | Percentile | Standard Error | Lower 99.0% Normal CI | Upper 99.0% Normal CI |
|---|---|---|---|---|---|
| 0.1 | 5 | 0.114565 | 0.0200486 | 0.0729939 | 0.179811 |

The percentile value shows the fatigue of human with the mentioned specifications. The effect of stress and available time on fatigue, are shown in *Figs. 3-4*. It could be seen, that fatigue increases when stress is increasing and available time is decreasing.

To verify the accuracy of the proposed model, five instances have been investigated, and the relative error has been calculated for each instance. The provided results shown in *Table 8* indicated the proposed model could estimate human fatigue with an acceptable error range. In other words, reducing the time and cost by using the PCA and ALT did not cause a significant error.

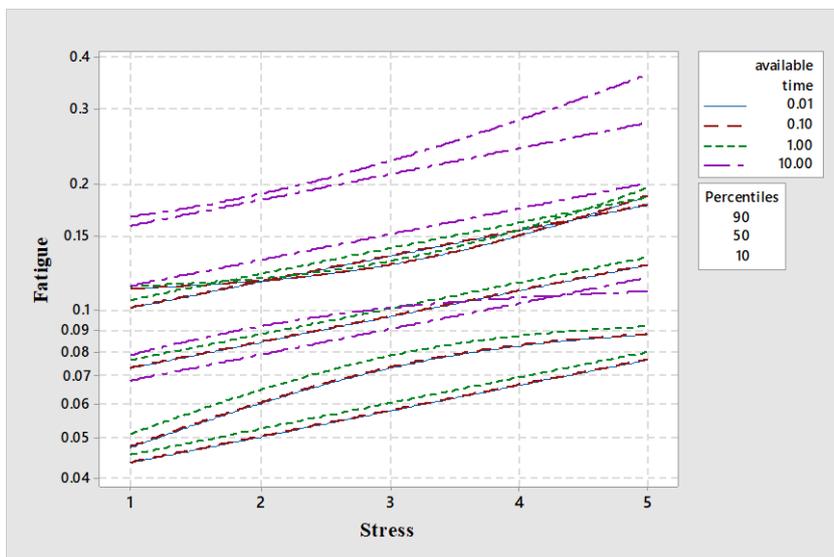

**Fig. 3.** The relation between fatigue and stress in different stress value.

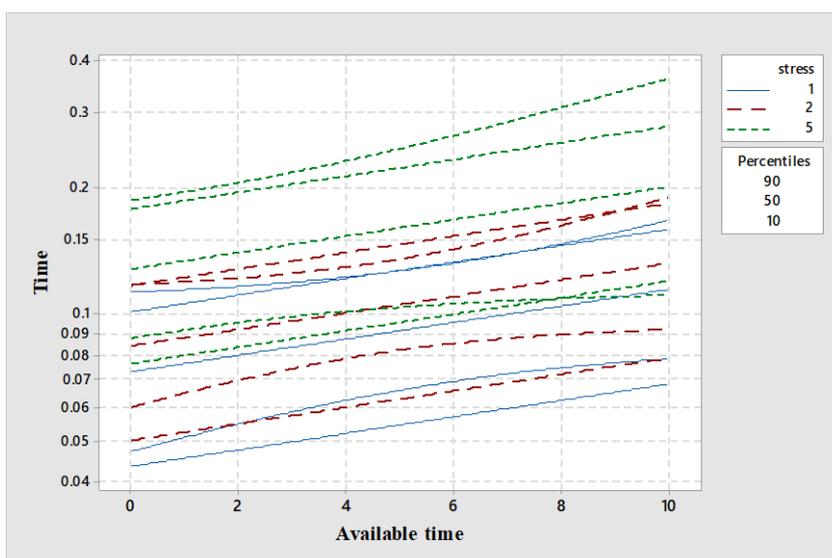

**Fig. 4.** The relation between fatigue and available time in different time value.

Table 8. The relative fatigue error.

| | Available Time | Stress | Complexity | Experience And Training | Procedures | Ergonomics | Fitness For Duty | Work Process | Fatigue | ALT- Fatigue | Relative Error |
|---|---|---|---|---|---|---|---|---|---|---|---|
| Ins 1 | 10 | 5 | 5 | 0.5 | 20 | 10 | 5 | 1 | 0.195 | 0.216 | 0.1060 |
| Ins2 | 1 | 5 | 2 | 0.5 | 50 | 0.5 | 1 | 5 | 0.062 | 0.069 | 0.1150 |
| Ins 3 | 1 | 5 | 1 | 0.5 | 50 | 10 | 1 | 1 | 0.073 | 0.080 | 0.0980 |
| Ins 4 | 10 | 5 | 1 | 1 | 5 | 1 | 5 | 5 | 0.162 | 0.175 | 0.0830 |
| Ins 5 | 0.01 | 5 | 2 | 1 | 1 | 10 | 5 | 1 | 0.114 | 0.130 | 0.1380 |

# 4 | Conclusion and Future Work

In every manufacturing systems, human has the most impact on quality and safety, regardless of the role of human, every plan such as production scheduling, quality improvement, and cost reduction are doomed to failure. Many methods have been proposed to quantify the human effects in manufacturing systems. Most of these methods require a lot of data collection and several experiments. Considering this fact, in this paper, we tried to propose a simple method for human resource fatigue calculation. In this paper, we used the ALT to estimate human fatigue using PSFs. The proposed model can decrease the data gathering time in comparison with common human fatigue models. To reduce the amount of required data for fatigue calculation, the PCA method has been implemented to find the most effective PSFs. The performance of the proposed method was examined for a real case (lathing manufacturing system), and the provided results indicated that the proposed model could obtain an efficient and effective value for human fatigue using fewer data and costs. It's worth mentioning that future researches could consider the potential relation of PSFs and another formulas of ALT. Investigating the effect of another formulas for ALT on human fatigue prediction error, can be an interesting issue for future works.